\documentclass[12pt]{article}
\usepackage{amsfonts}
\usepackage{amsmath}

\newcommand{\R} {{\rm I}\!{\rm R}}

\begin{document}

\title{\bf Two-dimensional integrable theories with \\ 
 defects and fibre bundles over the circle}
\author{\bf E. P. Gueuvoghlanian  \\ 
 Instituto de F\' \i sica Te\'orica, 
UNESP, \\ 
Rua Pamplona 145,  01405-900,  S\~ao Paulo, SP, Brazil. \\ 
e-mail:gueuvo@ift.unesp.br}
\date{}

\maketitle

\begin{abstract}
A procedure is described to associate fibre bundles over the circle to 
two-dimensional theories with defects which have their field equations and 
defects described by a zero curvature condition.
\end{abstract}

\setcounter{equation}{0}
\def\theequation{\thesection. \arabic{equation}}

\section{Introduction}

The lagrangian formulation of two-dimensional integrable theories with 
certain discontinuites called defects has been the subject of some works 
in the last years. The theories analyzed include the cases of the Liouville,
sine-Gordon and abelian affine Toda theories \cite{um}-\cite{dois}, 
the nonlinear Schr\"odinger theory \cite{tres}, the complex sine-Gordon 
theory \cite{quatro} and supersymmetric extensions of the sine-Gordon theory 
\cite{cinco}-\cite{seis}.

In the section 2 we review the Liouville theory with defect. In the section
3 we consider general theories with defects admiting a zero curvature 
description for their field equations and defects. We show that these field
theories define in a natural way fibre bundles over the circle in such a 
way that theories in the absence of the defect correspond to trivial fibre
bundles. In the appendixes A and B two calculations of the section 2 are 
given. 

\section{The Liouville theory with defect}

The Liouville theory is an abelian conformal Toda theory. Its field equation
is given in terms of a zero curvature condition
\begin{equation}
\partial\bar{A} - \bar{\partial} A +[A,\bar{A}]=0, \label{2.1}
\end{equation}
where
\begin{eqnarray}
&& A = B \varepsilon^-  B^{-1}, \label{2.2} \\
&& \bar{A}=- \varepsilon^+ - \bar{\partial} B B^{-1}, \label{2.3}
\end{eqnarray}
where $z \equiv t+x$, $\bar{z} \equiv t-x$,  
$\partial \equiv \frac{\partial}{\partial z} = (1/2) (\partial_t + \partial_x)
$, $\bar{\partial} \equiv \frac{\partial}{\partial \bar{z} } = (1/2) 
(\partial_t - \partial_x) $, $ A \equiv A_z = (1/2)(A_t+A_x)$, $\bar{A} \equiv
A_{\bar{z}} =(1/2)(A_t-A_x)$, $B \equiv \exp (\phi h)$, $\varepsilon^+ \equiv
\mu E_{\alpha}$, $\varepsilon ^- \equiv \mu E_{-\alpha}$, $\mu \in \R$ and 
$(h, E_{\alpha}, E_{-\alpha})$ are the Chevalley generators of the Lie 
algebra $A_1$:  
\begin{eqnarray}
&&[h,E_{\pm \alpha}]= \pm 2 E_{\pm \alpha} \ \ \hbox{and} \label{2.4} \\   
&&[E_{\alpha}, E_{-\alpha}]=h. \label{2.5}
\end{eqnarray}
The field equation is
\begin{equation}
\partial \bar{\partial} \phi = \mu^2 e^{-2\phi}. \label{2.6}
\end{equation}
(A parametrization $B \equiv \exp (-\phi h)$, $\varepsilon^+ \equiv
\mu E_{\alpha}$ and $\varepsilon ^- \equiv -\mu E_{-\alpha}$ would result
in the equation $\partial \bar{\partial} \phi = \mu^2 e^{2\phi}$.) 

There is a standard procedure \cite{sete} to obtain the lagrangian of the 
Toda theories using their relation to the Wess-Zumino-Witten theory. The 
lagrangian of the Liouville theory is
\begin{equation}
L=- \frac{k}{2\pi} \partial \phi \bar{\partial} \phi + 
\frac{k \mu^2}{2\pi} e^{-2\phi} , \label{2.7}
\end{equation}
where $k$ is constant. The theory defined over all the real line $\R$, that 
is, $ x \in \R$, in the absence of the defect, is denominated the theory in 
the bulk. The lagrangian of the Liouville theory with defect is given by
\begin{equation}
L= \theta(-x) L_1 + \theta(x) L_2 + \delta(x) L_D \label{2.8}
\end{equation}
where
\begin{eqnarray}
&&L_p = - \frac{k}{2\pi} \partial \phi_p \bar{\partial} \phi_p +
\frac{k \mu^2}{2\pi} e^{-2\phi_p}, \label{2.9} \\ 
&&L_D= \frac{k}{8\pi}(\phi_2 \partial_t \phi_1 -\phi_1 \partial_t \phi_2)
+ B(\phi_1,\phi_2), \label{2.10}
\end{eqnarray}
$ p \in \{ 1,2 \}$, $\delta$ is the Dirac delta function, $\theta$ is the step
function, $( d \theta (x) / dx) = \delta(x)$, with the choice $\theta (0) =0$,
that is, $\theta(x) =1$ if $x > 0$ and $\theta (x) =0$ if $x \leq 0$. Note that
this choice implies that
\begin{eqnarray}
&&\int^a_0 \delta(x) dx = \int^a_0 \frac{d\theta(x)}{dx} dx =1
=- \int^0_a \delta(y) dy= \int^0_{-a} \delta(-x) dx 
, \label{2.11} \\ 
&&\int^0_{-a} \delta(x) dx = \int^0_{-a} \frac{d\theta(x)}{dx} dx =0
=- \int^{-a}_0 \delta(y) dy= \int^a_0 \delta(-x) dx, \label{2.12} 
\end{eqnarray}
$ \forall a > 0$. The functional $B(\phi_1,\phi_2)$ is called the border function
and it depends on the fields $\phi_1$ and $\phi_2$ but it does not depend on any
of their derivatives. The border function is going to be determinated by an 
argument related to the conservation of a modified momentum assigned to the theory.
We have that
\begin{eqnarray}
\frac{\delta L}{\delta \phi_1} &=& - \frac{k\mu^2}{\pi} \theta(-x) e^{-2\phi_1}+
\delta(x) \left( \frac{\delta B}{\delta \phi_1} 
- \frac{k}{8 \pi} \partial_t \phi_2  \right) , \label{2.13} \\ 
\frac{\delta L}{\delta \phi_2} &=& - \frac{k\mu^2}{\pi} \theta(x) e^{-2\phi_2}+
\delta(x) \left( \frac{\delta B}{\delta \phi_2} 
+ \frac{k}{8 \pi} \partial_t \phi_1  \right) , \label{2.14} \\ 
\partial_{\mu} \frac{\delta L}{\delta (\partial_{\mu} \phi_1)} 
&=& - \frac{k}{4 \pi} \theta(-x) (\partial^2_t - \partial^2_x) \phi_1
- \frac{k}{4 \pi} \delta(-x) \partial_x \phi_1  
+ \frac{k}{8 \pi} \delta(x) \partial_t \phi_2, \label{2.15} \\ 
\partial_{\mu} \frac{\delta L}{\delta (\partial_{\mu} \phi_2)} 
&=& - \frac{k}{4 \pi} \theta(x) (\partial^2_t - \partial^2_x) \phi_2
 - \frac{k}{8 \pi} \delta(x) \partial_t \phi_1  
 + \frac{k}{4 \pi} \delta(x) \partial_x \phi_2. \label{2.16}     
\end{eqnarray}
The Euler-Lagrange equations
\begin{equation}
\frac{ \delta L}{\delta \phi_p } = \partial_{\mu} \left( 
\frac{\delta L} {\delta( \partial_{\mu} \phi_p )}  \right), \label{2.17}
\end{equation}
$ p \in \{ 1,2 \} $, are equivalent to:

a) If $ x>0 $:
\begin{equation}
\partial \bar{\partial} \phi_2 = \mu^2 e^{-2\phi_2}. \label{2.18}
\end{equation}

b) If $ x<0 $:
\begin{equation}
\partial \bar{\partial} \phi_1 = \mu^2 e^{-2\phi_1}. \label{2.19}
\end{equation}

c) If $x=0$:
\begin{eqnarray}
&&\partial_t \phi_2 - \partial_x \phi_1 
= \frac{4 \pi}{k} \frac{\delta B}{\delta \phi_1} \ \ \hbox{and} \label{2.20} \\
&&\partial_x \phi_2 - \partial_t \phi_1 
= \frac{4 \pi}{k} \frac{\delta B}{\delta \phi_2}. \label{2.21}
\end{eqnarray}
The equations (\ref{2.20}) and (\ref{2.21}) are called the defect conditions.
Note that we do not have a condition $\phi_1=\phi_2$ at $x=0$. 

The energy-momentum tensor is given by
\begin{equation}
T^{\mu\nu}= \sum^2_{p=1} g^{\nu\gamma} \frac{\delta L} {\delta( \partial_{\mu} \phi_p )}
\partial_{\gamma} \phi_p - g^{\mu\nu} L \label{2.22}
\end{equation}
where $g_{00}=g^{00}=1$, $g_{11}=g^{11}=-1$, $g_{01}=g_{10}=g^{01}=g^{10}=0$ and
$(t,x)=(x^0,x^1)$.
Thus
\begin{equation}
P=\int^0_{-\infty} T^{01} dx + \int^{\infty}_0 T^{01} dx 
 = \frac{k}{4\pi} \int^0_{-\infty} \frac{\partial \phi_1}{\partial t}
 \frac{\partial \phi_1}{\partial x} dx +
\frac{k}{4\pi} \int^{\infty}_{0} \frac{\partial \phi_2}{\partial t}
 \frac{\partial \phi_2}{\partial x} dx. \label{2.23} 
\end{equation}
The time derivative of $P$ is
\begin{eqnarray}
\frac{dP}{dt} \equiv \frac{ dP^{(1)} }{dt} + \frac{ dP^{(2)} }{dt} 
&=&\frac{k}{4\pi} \int^0_{-\infty}[(\partial_t\partial_x\phi_1)\partial_t\phi_1+
(\partial^2_t\phi_1)\partial_x\phi_1] dx \nonumber \\
&+& \frac{k}{4\pi} \int^{\infty}_{0}[(\partial_t\partial_x\phi_2)\partial_t\phi_2+
(\partial^2_t\phi_2)\partial_x\phi_2] dx. \label{2.24}
\end{eqnarray}
Using (\ref{2.19}), we have
\begin{equation}
 \frac{ dP^{(1)} }{dt} = \frac{k}{4\pi} \int^0_{-\infty} 
\left[ (\partial_t\partial_x\phi_1)\partial_t\phi_1
+ \partial_x \phi_1 \left( \partial^2_x\phi_1 +  \frac{4\pi}{k} 
\frac{ \delta V_1 }{\delta \phi_1} \right) \right ] dx, \label{2.25} 
\end{equation}
where
\begin{equation}
V_p \equiv - \frac{k\mu^2}{2\pi} e^{-2\phi_p}, \label{2.26}
\end{equation}
$p \in \{ 1,2 \}$. Then
\begin{equation}
\frac{ dP^{(1)} }{dt} = \frac{k}{4\pi} \int^0_{-\infty}
\partial_x \left[ \frac{(\partial_t\phi_1)^2}{2} +
\frac{(\partial_x\phi_1)^2}{2} + \frac{4\pi}{k} V_1 \right] dx. \label{2.27}
\end{equation}
The other term in (\ref{2.24}) can be treated in a similar way. Thus
\begin{equation}
\frac{ dP }{dt}=
\frac{k}{4\pi} \left[ \frac{(\partial_t\phi_1)^2}{2} +
\frac{(\partial_x\phi_1)^2}{2} + \frac{4\pi}{k} V_1 \right]  \big{|}  ^0_{-\infty}
+ \frac{k}{4\pi} \left[ \frac{(\partial_t\phi_2)^2}{2} +
\frac{(\partial_x\phi_2)^2}{2} + \frac{4\pi}{k} V_2 \right]  \big{|}  ^{\infty}_0. 
\label{2.28}  
\end{equation}
The kind of terms at $ \pm \infty $ are already present in theory defined in the
bulk (in the absence of the defect) and we are going to ignore them. Considering 
only the contribution from $x=0$ and using the expressions for $\partial_x \phi_1$ 
and $\partial_x \phi_2$ from (\ref{2.20}) and (\ref{2.21}), we have
\begin{equation}
\frac{ dP }{dt}= \left[ -\partial_t\phi_2 \frac{\delta B}{\delta \phi_1}
 -\partial_t\phi_1 \frac{\delta B}{\delta \phi_2}
+ \frac{2\pi}{k} \left( \frac{\delta B}{\delta \phi_1} \right) ^2
- \frac{2\pi}{k} \left( \frac{\delta B}{\delta \phi_2} \right) ^2
+V_1-V_2 \right] \big{|}_{x=0}. \label{2.29}
\end{equation}
If
\begin{eqnarray}
&&\left[ \frac{2\pi}{k} \left( \frac{\delta B}{\delta \phi_1} \right) ^2
     - \frac{2\pi}{k} \left( \frac{\delta B}{\delta \phi_2} \right) ^2
     +V_1-V_2 \right] =0,  \label{2.30} \\
&&\frac{\delta B (\phi_1,\phi_2)}{\delta \phi_1}=
\frac{\delta M (\phi_1,\phi_2)}{\delta \phi_2} , \label{2.31} \\
&&\frac{\delta B (\phi_1,\phi_2)}{\delta \phi_2}=
\frac{\delta M (\phi_1,\phi_2)}{\delta \phi_1} , \label{2.32} 
\end{eqnarray}
for some functional $M$, in such a way that 
\begin{equation}
\frac{ \delta^2 B }{\delta \phi^2_1} = 
\frac{ \delta^2 B }{\delta \phi^2_2}, \label{2.33} 
\end{equation}
then
\begin{equation}
\left( \frac{dP}{dt} \right) \big{|} _t =
- \left( \frac{\partial}{\partial t} \left[ M(t,x) \right] \right) \big{|} _{(t,x=0)} =
- \left( \frac{d}{d t} \left[ M(t,x=0) \right] \right) \big{|} _t \label{2.34}
\end{equation}
and
\begin{equation}
\frac{d}{d t} \left[ P+M(t,x=0) \right] =0. \label{2.35}
\end{equation}
Thus the modified momentum $P+M(t,x=0)$ is conserved provided
we have the relations (\ref{2.30})-(\ref{2.33}).

The energy is given by
\begin{eqnarray}
E&=& \int^0_{-\infty} T^{00} dx + \int^{\infty}_0 T^{00} dx \nonumber \\
&=& \frac{k}{8\pi} \int^0_{-\infty} \left[ 
\frac{8\pi }{k} V_1 
-( \partial_t \phi_1)^2
-( \partial_x \phi_1)^2
  \right] dx \nonumber \\
&+&\frac{k}{8\pi} \int^{\infty}_0 \left[ 
 \frac{8\pi }{k} V_2
-( \partial_t \phi_2)^2
-( \partial_x \phi_2)^2
  \right] dx. \label{2.36} 
\end{eqnarray}
Using (\ref{2.18}) and (\ref{2.19}) the time derivative of $E$ can be evaluated as
\begin{equation}
\frac{dE}{dt} = - \frac{k}{4\pi} ( \partial_t \phi_1 \partial_x \phi_1 ) |^0_{-\infty}
- \frac{k}{4\pi} ( \partial_t \phi_2 \partial_x \phi_2 ) |^{\infty}_{0}. \label{2.37}
\end{equation}
Ignoring the terms at $\pm \infty$ and using the expressions for $\partial_x\phi_1$ 
and  $\partial_x\phi_2$ from (\ref{2.20}) and (\ref{2.21}), we have:
\begin{equation}
\left( \frac{dE}{dt} \right) \big{|} _t =
 \left( \frac{\partial}{\partial t} \left[ B(t,x) \right] \right) \big{|} _{(t,x=0)} =
 \left( \frac{d}{d t} \left[ B(t,x=0) \right] \right) \big{|} _t \label{2.38}
\end{equation}
and
\begin{equation}
\frac{d}{d t} \left[ E-B(t,x=0) \right] =0. \label{2.39}
\end{equation}
Thus the modified energy $ E-B(t,x=0)$ is conserved. Note that we do not
need the expressions (\ref{2.30})-(\ref{2.33}) to get (\ref{2.39}). We introduce
$ \phi^+ \equiv \phi_1+\phi_2 $ and $\phi^- \equiv \phi_1 -\phi_2$. Then 
\begin{eqnarray}
\frac{ \delta}{\delta \phi^+ } &=& \frac{1}{2} 
\left( \frac{ \delta}{\delta \phi_1 } + \frac{ \delta}{\delta \phi_2 }
\right) \ \ \hbox{and} \label{2.40} \\
\frac{ \delta}{\delta \phi^- } &=& \frac{1}{2} 
\left( \frac{ \delta}{\delta \phi_1 } - \frac{ \delta}{\delta \phi_2 }
\right) . \label{2.41}
\end{eqnarray}
If we write 
\begin{equation}
B(\phi_1,\phi_2) \equiv B^+(\phi^+) + B^-(\phi^-), \label{2.42}
\end{equation}
we see that
(\ref{2.31})-(\ref{2.32}) is solved by 
\begin{equation}
M= B^+ - B^-. \label{2.43}
\end{equation}
Using (\ref{2.40})-(\ref{2.42}), we see that the equation (\ref{2.30}) is 
equivalent to 
\begin{equation}
2 \frac{\delta B^+}{\delta \phi^+} \frac{\delta B^-}{\delta \phi^-}
=-  \left( \frac{k}{2\pi} \right) ^2 \mu^2 e^{-\phi^+} \sinh (\phi^-).
\label{2.44}
\end{equation}
This equation is solved by
\begin{eqnarray}
&&B^+ = \frac{k}{2\pi} \mu \lambda e^{-\phi^+} \ \ \hbox{and} \label{2.45} \\
&&B^- = \frac{1}{2} \frac{k}{2\pi} \frac{\mu}{\lambda} \cosh{\phi^-}, \label{2.46}
\end{eqnarray}
where $\lambda$ is an arbitrary constant.

The defect conditions, given by the equations (\ref{2.20})-(\ref{2.21}), are
equivalent to
\begin{eqnarray}
&&\partial( \phi_1-\phi_2) |_{x=0} = 2 \mu \lambda e^{-\phi^+} |_{x=0}
\ \ \hbox{and} \label{2.47} \\
&&\bar{ \partial } ( \phi_1+\phi_2) |_{x=0} = \frac{\mu}{\lambda} \sinh(\phi^-) |_{x=0} .
\label{2.48}
\end{eqnarray}
Suppose that the equations (\ref{2.47}) and (\ref{2.48}) hold not only at 
$(t,x=0)$ but at all $(t,x)$. Then taking the $\bar{\partial}$ derivative of
(\ref{2.47}) and using (\ref{2.48}), results
\begin{equation}
\bar{\partial} \partial(\phi_1-\phi_2) = \mu^2(e^{-2\phi_1}-e^{-2\phi_2}).
\label{2.49}
\end{equation}
Similarly, taking the $\partial$ derivative of (\ref{2.48}) and using (\ref{2.47}),
results
\begin{equation}
\bar{\partial} \partial(\phi_1+\phi_2) = \mu^2(e^{-2\phi_1}+e^{-2\phi_2}).
\label{2.50}
\end{equation}
We see that (\ref{2.49}) and (\ref{2.50}) imply that ($\phi_1$ is a solution
of the Liouville equation) $\Longleftrightarrow$ ($\phi_2$ is a solution
of the Liouville equation). That is, if the equations (\ref{2.47}) and
(\ref{2.48}) were valid at all the points $(t,x)$ they would be B\"{a}cklund
transformations. As (\ref{2.47}) and (\ref{2.48}) are supposed to hold only
at the point $(t,x=0)$, they are said to be frozen B\"{a}cklund transformations.

From (\ref{2.2}) and (\ref{2.3}) we can obtain $A_t$ and $A_x$. It is 
convenient to us to apply a gauge transformation to $(A_t,A_x)$ with a group
element given by $g=e^{ \frac{\phi h}{2} }$ to obtain $(A^g_t,A^g_x)$. To
simplify our notation we are just going to write $(A_t,A_x)$ instead 
$(A^g_t,A^g_x)$:
\begin{eqnarray}
&&A_t= -\mu e^{-\phi} E_{\alpha} + \mu e^{-\phi} E_{-\alpha}
+ \frac{\partial_x \phi h}{2} \ \ \hbox{and} \label{2.51} \\
&&A_x= \mu e^{-\phi} E_{\alpha} + \mu e^{-\phi} E_{-\alpha}
+ \frac{\partial_t \phi h}{2}.  \label{2.52}
\end{eqnarray}
As the curvature transforms as $F^g_{tx}=gF_{tx}g^{-1}$, we see that
($F^g_{tx}=0$) $\Longleftrightarrow$ ($F_{tx}=0$). That is, the zero 
curvature condition
\begin{equation}
\partial_t A_x - \partial_x A_t +[A_t,A_x]=0 \label{2.53}
\end{equation}
is equivalent to the Liouville equation (\ref{2.6}), where $A_t$ and
$A_x$ are given by (\ref{2.51}) and (\ref{2.52}).

Now we want to describe a procedure \cite{um} to obtain the field 
equations (\ref{2.18}) - (\ref{2.19}) and the defect conditions
(\ref{2.20}) - (\ref{2.21}) as a zero curvature condition in the
following sense: Define
\begin{eqnarray}
A^{(p)}_t &\equiv&  -\mu e^{-\phi_p} E_{\alpha} + \mu e^{-\phi_p} E_{-\alpha}
+ \frac{\partial_x \phi_p h}{2}, \label{2.54} \\
A^{(p)}_x &\equiv&  \mu e^{-\phi_p} E_{\alpha} + \mu e^{-\phi_p} E_{-\alpha}
+ \frac{\partial_t \phi_p h}{2}, \label{2.55} \\
D_1 &\equiv& \partial_x \phi_1 - \partial_t \phi_2  
+ \frac{4 \pi}{k} \frac{\delta B}{\delta \phi_1} \ \ \hbox{and} \label{2.56} \\
D_2 &\equiv& \partial_x \phi_2 - \partial_t \phi_1  
- \frac{4 \pi}{k} \frac{\delta B}{\delta \phi_2}, \label{2.57}
\end{eqnarray}
$p \in \{ 1,2 \} $. Note that $D_1=0$ and $D_2=0$ are equivalent to
(\ref{2.20}) and (\ref{2.21}). 

Let $a,b \in \R$, $a<0<b$. Define

a) $\forall (t,x) \in \R ^2$ such that $x<b$:
\begin{eqnarray}
&&\hat{A}^{(1)}_t \equiv [ \theta(x-a)+ \theta(a-x)] A^{(1)}_t 
- \frac{1}{2} \theta(x-a) D_1 h \ \ \hbox{and} \label{2.58} \\
&&\hat{A}^{(1)}_x \equiv \theta(a-x) A^{(1)}_x. \label{2.59}
\end{eqnarray}

b) $\forall (t,x) \in \R ^2$ such that $x>a$:
\begin{eqnarray}
&&\hat{A}^{(2)}_t \equiv [ \theta(x-b)+ \theta(b-x)] A^{(2)}_t 
- \frac{1}{2} \theta(b-x) D_2 h \ \ \hbox{and} \label{2.60} \\
&&\hat{A}^{(2)}_x \equiv \theta(x-b) A^{(2)}_x. \label{2.61}
\end{eqnarray} 

Let 
\begin{eqnarray}
&&\partial_t \hat{A}^{(1)}_x - \partial_x \hat{A}^{(1)}_t 
+[\hat{A}^{(1)}_t,\hat{A}^{(1)}_x]=0 
\ \ \hbox{and} \label{2.62} \\
&&\partial_t \hat{A}^{(2)}_x - \partial_x \hat{A}^{(2)}_t 
+[\hat{A}^{(2)}_t,\hat{A}^{(2)}_x]=0. 
\label{2.63}
\end{eqnarray}
One can verify, as explained in the appendix A, that:

a) The equation (\ref{2.62}) implies for $x<a$ in the 
equation (\ref{2.53}) calculated at $(A_t,A_x) = (A^{(1)}_t,A^{(1)}_x)$ and
this is equivalent to (\ref{2.19}).

b) The equation (\ref{2.63}) implies for $x>b$ in the 
equation (\ref{2.53}) calculated at $(A_t,A_x) = (A^{(2)}_t,A^{(2)}_x)$ and
this is equivalent to (\ref{2.18}).

c) The equation (\ref{2.62}) implies for $x=a$ in $D_1=0$ and this is 
equivalent to \\ (\ref{2.20}).

d) The equation (\ref{2.63}) implies for $x=b$ in $D_2=0$ and this is
equivalent to \\ (\ref{2.21}).

e) In the intersection $a<x<b$, the equations (\ref{2.62}) and (\ref{2.63})
imply that 
\begin{equation}
\partial_x \phi_1 = \partial_x \phi_2 =0. \label{2.64}
\end{equation}
The idea is that we can take $|a|$ and $b$ so small as we want and in
this case the conditions a) to d) would correspond to the field equations
and defect conditions as expressed in (\ref{2.18}) - (\ref{2.21}).

Note that the intersection of the domains in which 
$(\hat{A}^{(1)}_t,\hat{A}^{(1)}_x)$ and $(\hat{A}^{(2)}_t,\hat{A}^{(2)}_x)$
are defined is $(t,x) \in \R ^2$ such that $a<x<b$. Suppose that in this
intersection the two connections are related by a gauge transformation 
\cite{um}. In this case, \\ (\ref{2.62}) $\Longleftrightarrow$ (\ref{2.63}).
Note that in the intersection $a<x<b$ we have \\ 
$\hat{A}^{(1)}_x=\hat{A}^{(2)}_x=0$. Then the group element $g$ that gives the
gauge transformation must be $x$-independent. We need to find $g$ such that
\begin{equation}
\hat{A}^{(1)}_t= g \hat{A}^{(2)}_t g^{-1} - (\partial_t g)g^{-1}. \label{2.65}
\end{equation}
We can verify, as explained in the appendix B, that
\begin{equation}
g= e^{- \frac{\phi_2 h}{2} } e^{2 \lambda E_{\alpha} } 
e^{ \frac{\phi_1 h}{2} }  \label{2.66}
\end{equation}
where $\lambda$ is the constant introduced in the equations (\ref{2.45}) - 
(\ref{2.46}). Note that $g$ is $x$-independent as a consequence of (\ref{2.64})
but it can be $t$-dependent.

\section{Two-dimensional integrable theories with \\ 
 defects and fibre bundles over the circle}

\setcounter{equation}{0}
\def\theequation{\thesection. \arabic{equation}}
 
In this section we consider general theories which have the following
structure: their lagrangians have the form
\begin{equation}
L= \theta(-x) L_1 + \theta(x) L_2 + \delta(x) L_D. \label{3.1}
\end{equation}
The field equations and defect conditions corresponding to (\ref{3.1})
can be expressed as a zero curvature condition associated to two sets
of gauge potentials. That is, it is possible to define

a)$\hat{A}^{(1)}_t$ and $\hat{A}^{(1)}_x$, $\forall (t,x) \in \R ^2$ such
that $x<b$;

b)$\hat{A}^{(2)}_t$ and $\hat{A}^{(2)}_x$, $\forall (t,x) \in \R ^2$ such
that $x>a$; where $a<0<b$, such that
\begin{eqnarray}
&&\partial_t \hat{A}^{(1)}_x - \partial_x \hat{A}^{(1)}_t 
+[\hat{A}^{(1)}_t,\hat{A}^{(1)}_x]=0 
\ \ \hbox{and} \label{3.2} \\
&&\partial_t \hat{A}^{(2)}_x - \partial_x \hat{A}^{(2)}_t 
+[\hat{A}^{(2)}_t,\hat{A}^{(2)}_x]=0 
\label{3.3}
\end{eqnarray}
correspond to the field equations (for $x<a$ and $x>b$) and defect 
conditions (for $x=a$ and $x=b$) in an analogous way to that described
in the section 2 in the case of the Liouville theory with defect. In the
intersection, $(t,x) \in \R ^2$ such that $a<x<b$, (\ref{3.2}) and 
(\ref{3.3}) imply that the fields of the theory obey some additional set of
equations, as (\ref{2.64}) in the case of the Liouville theory with defect.
We suppose that in this intersection the two sets of gauge potentials are
related by a gauge transformation
\begin{eqnarray}
\hat{A}^{(1)}_t= g \hat{A}^{(2)}_t g^{-1} - (\partial_t g)g^{-1}, \label{3.4} \\
\hat{A}^{(1)}_x= g \hat{A}^{(2)}_x g^{-1} - (\partial_x g)g^{-1}, \label{3.5}
\end{eqnarray}
where $g$ is a group element.

That is the situation we are proposing to analyze. The key role is played by the 
group element $g$. We suppose that the defect is placed at $x=0$, but this can
be generalized easely. As $a$ and $b$ are arbitrary, to simplify our discussion,
we are going to take $ |a|=b=r>0 $.

Note that in the Liouville theory with defect we were in fact dealing with the 
Minkowski $Mk^2$ space- see (\ref{2.22}) -. That is, the differentiable manifold
$ \R ^2 $ with the metric tensor $g^{metric}=dt \otimes dt -dx \otimes dx$ (in 
global coordinates) and a Levi-Civita connection. In this section we consider
the differentiable manifold $\R ^2$ with the constant metric tensor given by
the general expression 
$g^{metric}= \epsilon_t dt \otimes dt + \epsilon_x dx \otimes dx$, where 
$\epsilon_t= \pm 1$ and $\epsilon_x= \pm 1$ (in global coordinates) and a Levi-Civita
connection. The Euclidean space $E^2$ corresponds to $\epsilon_t=  1$ and
$\epsilon_x=  1$.

Now we are going to review
the definition of fibre bundles. Then we are
going to see how a fibre bundle can be constructed from a minimal information and 
apply that process to our case.

A coordinate bundle \cite{Nakahara} $(E,\pi, M, F, G, \{ U_i \} , \{ \phi_i \} )$
consists of the following elements:

a) A differentiable manifold $E$ called the total space; a
differentiable manifold $M$ called the base space; 
a differentiable manifold $F$ called the fibre.
  
b) A surjective map $\pi:E \rightarrow M$ called the projection.
The inverse image $\pi^{-1} \{ p \} \equiv F_p$ is called the fibre at $p$,
$ \forall p \in M$.

c) A Lie group $G$ called the structure group, which acts on $F$ on the
left. That is, there is a differentiable map $T:G \times F \rightarrow F$ such that
$T(g_1,T(g_2,f))=T(g_1g_2,f)$ and $T(e,f)=f$, $\forall g_1,g_2 \in G$,
$\forall f \in F$, where $e$ is the identity element of $G$. (Note that
a Lie group is, by definition, a differentiable manifold.) We simplify
the notation writting $T(g,f) \equiv gf$. 

d) A set $ \{ U_i \} $ which is an open covering of $M$ $( \bigcup _i U_i =M)$ 
and a set $ \{ \phi_i \} $, where each $\phi_i$ is a diffeomorphism 
$\phi_i:U_i \times F \rightarrow \pi^{-1} (U_i)$ such that $\pi \phi_i (p,f)=p$,
$\forall p \in U_i$, $\forall f \in F$. The map $\phi_i$ is called the local 
trivialisation.

e) We define $\phi_{i,p} (f) \equiv \phi_i (p,f)$, $\forall p \in U_i$,
$\forall f \in F$. Then the map $ \phi_{i,p}:F \rightarrow F_p$ is a 
diffeomorphism. If $U_i \bigcap U_j \neq \emptyset$, we require that
$t_{ij}(p) \equiv \phi_{i,p}^{-1} \phi_{j,p} : F \rightarrow F$ be an
element of $G$. Then $\phi_i$ and $\phi_j$ are related by a $C^{\infty}$
differentiable map \\ $t_{ij}: U_i \bigcap U_j \rightarrow G$ as:
\begin{equation}
\phi_j(p,f_j)=\phi_i(p,t_{ij}(p)f_j). \label{3.6} 
\end{equation}
$\{ t_{ij} \}$ are denominated the transition functions.

Two coordinate bundles \\ $(E,\pi, M, F, G, \{ U_i \} , \{ \phi_i \} )$ and
 $(E,\pi, M, F, G, \{ V_i \} , \{ \psi_i \} )$ are said to be equivalent
if \\ $(E,\pi, M, F, G, \{ U_i \} \bigcup  \{ V_i \}  , \{ \phi_i \} \bigcup 
\{ \psi_i \} )$ is again a coordinate bundle. The fibre bundle is defined
as an equivalence class of coordinate bundles and is usually denoted by 
$(E,\pi, M, F, G)$. 

We require the consistency conditions:

a) $\forall p \in U_i$
\begin{equation}
t_{ii} (p) = e. \label{3.7}
\end{equation}

b) $\forall p \in U_i \bigcap U_j$
\begin{equation}
t_{ij}(p) = [t_{ji}(p)]^{-1}. \label{3.8}
\end{equation}

c) $\forall p \in U_i \bigcap U_j \bigcap U_k$
\begin{equation}
t_{ij}(p)t_{jk}(p)=t_{ik}(p). \label{3.9}
\end{equation}

If all the transition functions can be taken to be identity maps, the fibre
bundle is called trivial. A trivial bundle is a product manifold $M \times F$.

The minimal information to construct a fiber bundle \cite{Nakahara} is given
by \\ $ \{ M,F,G, \{ U_i \} , \{ t_{ij} \} \} $, where $M$ and $F$ are differentiable
manifolds, $G$ is a Lie group acting on $F$ by the left, $ \{ U_i \} $ is an 
open covering of $M$ and the set $ \{ t_{ij} \} $ is such that each 
$t_{ij}: U_i \bigcap U_j \rightarrow G$ is a $C^{\infty}$ differentiable map 
satisfying (\ref{3.7})-(\ref{3.9}).

We define
\begin{equation}
E_0 = \bigcup _i U_i \times F \label{3.10}
\end{equation}
and introduce an equivalence relation 
$\sim$ between $(p,f) \in U_i \times F$ and  $(q,f^{'}) \in U_j \times F$
if and only if $p=q$ and $f^{'} =t_{ji}(p)f$.

The total space is defined by
\begin{equation}
E \equiv E_0 / \sim . \label{3.11}
\end{equation}

Let us denote the equivalence class of $(p,f)$ by $[(p,f)]$.
Then  $[(p,f)] \in E$. The projection is given by
\begin{equation}
\pi: [(p,f)] \rightarrow p \label{3.12}
\end{equation}
and the local trivialisation is given by
\begin{equation}
\phi_i:(p,f) \rightarrow [(p,f)]. \label{3.13}
\end{equation}

It is possible to verify that this construction defines \cite{Nakahara}
a coordinate bundle. Then the fibre bundle is the equivalence class 
containing this coordinate bundle. 

Now let us apply this process to our case. We need to specify $M, F, G, 
\{ U_i \} , \{ t_{ij} \} $. We take $F$ as the Lie group itself. This
corresponds to consider a principal fibre bundle. Of course, having
a principal fibre bundle, one can construct its associated vector 
bundles \cite{Nakahara}.

Let the set $S^1(r)$ be defined by
\begin{equation}
S^1(r) \equiv \{ (t,x) \in \R ^2 | t^2+x^2=r^2 \}, \label{3.14}
\end{equation}
where $r \in \R$, $r>0$. We want to make $S^1(r)$ a topological space.
The open ball of radius $R$ ($R \in \R$, $R>0$) and center $(t_0,x_0) \in
\R ^2$ is defined as
\begin{equation}
B^R(t_0,x_0) \equiv \{ (t,x) \in \R ^2 |(t-t_0)^2+(x-x_0)^2<R^2 \} .
\label{3.15}
\end{equation}
The usual topology of $\R ^2$ is the one such that $V \subseteq \R ^2$ is
open if and only if, $\forall p \in V$, $\exists B^R(p) \subset V$ for some
radius $R>0$. Note that, in particular, $B^R(p)$ and $\R ^2- \{ p \}$ are 
open sets of $\R ^2$, $\forall p \in \R ^2$, $\forall R >0$. We give to 
$S^1(r)$ the topology induced by $\R ^2$. That is, $U \subseteq S^1(r)$
is open if and only if $U=S^1(r) \bigcap V$, where $V$ is some open set
of $\R ^2$. Note that, in particular, $S^1(r)- \{ p \}$ is an open set
of $S^1(r)$, $\forall p \in S^1(r)$, because
\begin{equation}
S^1(r) - \{ p \} = S^1(r) \bigcap \{ \R ^2 - \{ p \} \}. \label{3.16}
\end{equation}

It is possible to give to the topological space $S^1(r)$ a differentiable
structure in such a way it becomes a differentiable manifold \cite{Nakahara}.
We take $M$ as the differentiable manifold $S^1(r)$.

We need to specify an open covering $\{ U_i \}$ of $S^1(r)$. We define
\begin{eqnarray}
&&U_1 \equiv S^1(r) - \{ (t,x)=(0,-r) \} \ \ \hbox{and} \label{3.17} \\
&&U_2 \equiv S^1(r) - \{ (t,x)=(0,r) \} . \label{3.18}
\end{eqnarray}
Note that $(t,x)=(0,-r) \in S^1(r)$ and $(t,x)=(0,r) \in S^1(r)$.
Then $S^1(r)=U_1 \bigcup U_2$. Note that
\begin{equation}
U_1 \bigcap U_2 = A \bigcup B \label{3.19}
\end{equation}
where
\begin{eqnarray}
&&A \equiv \{ (t,x) \in S^1(r) | t>0 \}, \label{3.20} \\
&&B \equiv \{ (t,x) \in S^1(r) | t<0 \} \ \ \hbox{and} \label{3.21} \\
&&A \bigcap B = \emptyset. \label{3.22}
\end{eqnarray}

We need to specify $ \{ t_{ij} \}$. As our open covering has only
two open sets, we need to specify $C^{\infty}$ differentiable maps
$t_{12}:U_1 \bigcap U_2 \rightarrow G$, 
$t_{21}:U_1 \bigcap U_2 \rightarrow G$ 
and we do not need the equation (\ref{3.9}). We take
\begin{equation}
t_{21}(p) \equiv [t_{12}(p)]^{-1}, \label{3.23}
\end{equation}
in order to satisfy (\ref{3.8}).

As we said before, we are taking $|a|=b=r>0$. Let 
\begin{equation}
I= \{(t,x) \in \R ^2 | -r <x <r \}. \label{3.24}
\end{equation}
Then the group element $g$ in the equations (\ref{3.4}) and
(\ref{3.5}) defines a map $g:I \rightarrow G$. Note that
\begin{equation}
U_1 \bigcap U_2 = A \bigcup B \subset I. \label{3.25}
\end{equation}

Now consider the restriction of $g$ to $A \bigcup B$. If
$g:A \bigcup B \rightarrow G$ is a $ C^{\infty}$ differentiable
map, we can define $t_{12}: U_1 \bigcap U_2 \rightarrow G$ by
\begin{equation}
t_{12}(p) \equiv g(p), \label{3.26}
\end{equation}
$\forall p \in U_1 \bigcap U_2 = A \bigcup B$.  

By last, suppose we have a theory without defect such that
its field equations are expressed as a zero curvature 
condition in terms of the gauge potentials $A_t$ and $A_x$.
We can define

a)$\forall (t,x) \in \R ^2$ such that $x<b$
\begin{eqnarray}
&& \hat{A}^{(1)}_t(t,x)=A_t(t,x) \ \ \hbox{and} \label{3.27} \\
&& \hat{A}^{(1)}_x(t,x)=A_x(t,x). \label{3.28} 
\end{eqnarray}

b)$\forall (t,x) \in \R ^2$ such that $x>a$
\begin{eqnarray}
&& \hat{A}^{(2)}_t(t,x)=A_t(t,x) \ \ \hbox{and} \label{3.29} \\
&& \hat{A}^{(2)}_x(t,x)=A_x(t,x). \label{3.30} 
\end{eqnarray}
Where $a<0<b$. Then (\ref{3.2}) and (\ref{3.3}) correspond to the 
field equations without defect and in the intersection 
$ \{ (t,x) \in \R ^2 | a<x<b \} $ the gauge potentials are
related by a gauge transformation (\ref{3.4}) and (\ref{3.5}) where
$g$ is the identity element of $G$. It follows from (\ref{3.26}) that
all the transition functions are the identity element. Thus we have a
trivial fibre bundle.

\section{Appendix A}
 
\setcounter{equation}{0}
\def\theequation{\thesection. \arabic{equation}}

Using (\ref{2.58}) and (\ref{2.59}) we see that, $\forall (t,x) \in
\R ^2$ such that $x<b$, (\ref{2.62}) is equivalent to:
\begin{eqnarray}
&&\theta(a-x) \partial_t A^{(1)}_x 
-[\delta(x-a)-\delta(a-x)]A^{(1)}_t
-[\theta(x-a)+\theta(a-x)] \partial_x A^{(1)}_t \nonumber \\
&&+ \frac{1}{2} \theta(x-a) \partial_x D_1 h
+ \frac{1}{2} \delta(x-a) D_1h
-\frac{1}{2} \theta(x-a)\theta(a-x)D_1[h,A^{(1)}_x]  \nonumber \\
&&+[\theta(x-a)+\theta(a-x)]\theta(a-x)[A^{(1)}_t,A^{(1)}_x]=0. \label{4.1}
\end{eqnarray}

Similarly, using (\ref{2.60}) and (\ref{2.61}) we see that, 
$\forall (t,x) \in \R ^2$ such that $x>a$, (\ref{2.63}) is equivalent to:
\begin{eqnarray}
&&\theta(x-b) \partial_t A^{(2)}_x 
-[\delta(x-b)-\delta(b-x)]A^{(2)}_t
-[\theta(x-b)+\theta(b-x)] \partial_x A^{(2)}_t \nonumber \\
&&+ \frac{1}{2} \theta(b-x) \partial_x D_2 h
- \frac{1}{2} \delta(b-x) D_2h
-\frac{1}{2} \theta(x-b)\theta(b-x)D_2[h,A^{(2)}_x]  \nonumber \\
&&+[\theta(x-b)+\theta(b-x)]\theta(x-b)[A^{(2)}_t,A^{(2)}_x]=0. \label{4.2}
\end{eqnarray}

a) Let $x<a$. Then, from (\ref{4.1}),
\begin{equation}
\partial_t \hat{A}^{(1)}_x - \partial_x \hat{A}^{(1)}_t 
+[\hat{A}^{(1)}_t,\hat{A}^{(1)}_x]=0 . \label{4.3}
\end{equation}

b) Let $x>b$. Then, from (\ref{4.2}),
\begin{equation}
\partial_t \hat{A}^{(2)}_x - \partial_x \hat{A}^{(2)}_t 
+[\hat{A}^{(2)}_t,\hat{A}^{(2)}_x]=0 . \label{4.4}
\end{equation}

c)Let $x=a$. Then, from (\ref{4.1}),
\begin{equation}
D_1=0. \label{4.5}
\end{equation}

d)Let $x=b$. Then, from (\ref{4.2}),
\begin{equation}
D_2=0. \label{4.6}
\end{equation}

e)Let $a<x<b$. Then, from (\ref{4.1}) and (\ref{4.2}),
\begin{eqnarray}
&&-\partial_x A^{(1)}_t + \frac{1}{2}(\partial_xD_1)h=0 \ \ \hbox{and} \label{4.7} \\
&&-\partial_x A^{(2)}_t + \frac{1}{2}(\partial_xD_2)h=0. \label{4.8}
\end{eqnarray}
Using (\ref{2.42}), (\ref{2.45}), (\ref{2.46}) and (\ref{2.54})-(\ref{2.57}),  
we see that (\ref{4.7}) and (\ref{4.8}) are equivalent to
\begin{eqnarray}
&&-\mu \partial_x \phi_1 e^{-\phi_1}E_{\alpha}
+\mu \partial_x \phi_1 e^{-\phi_1}E_{-\alpha} \nonumber \\
&&+\left[ -\frac{\partial_x\partial_t\phi_2}{2}
 -\mu \lambda \partial_x(e^{-\phi_1-\phi_2})
+\frac{\mu}{2\lambda} \partial_x [\sinh(\phi_1-\phi_2)] \right] h=0
\ \ \hbox{and} \label{4.9} \\
&&-\mu \partial_x \phi_2 e^{-\phi_2}E_{\alpha}
+\mu \partial_x \phi_2 e^{-\phi_2}E_{-\alpha} \nonumber \\
&&+\left[ -\frac{\partial_x\partial_t\phi_1}{2} 
+\mu \lambda \partial_x(e^{-\phi_1-\phi_2})
+\frac{\mu}{2\lambda} \partial_x [\sinh(\phi_1-\phi_2)] \right] h=0. \label{4.10}
\end{eqnarray}
From (\ref{4.9}) and (\ref{4.10}), we see that
\begin{equation}
\partial_x \phi_1= \partial_x \phi_2 =0. \label{4.11}
\end{equation}

\section{Appendix B}
 
\setcounter{equation}{0}
\def\theequation{\thesection. \arabic{equation}}

We want to find the group element $g$ in such a way to satisfy the equation
(\ref{2.65}). In order to get an expression which has no derivatives we write
\begin{equation}
g=e^{-\frac{\phi_2 h }{2} } g_c e^{\frac{\phi_1 h }{2} }, \label{5.1}
\end{equation}
where $g_c$ is a group element which is t-independent and x-independent. After some
manipulation, we find that (\ref{2.65}) is equivalent to
\begin{eqnarray}
&&-(\mu e^{\phi_1-\phi_2})g_cE_{\alpha}g_c^{-1}
-(\mu e^{-\phi_1-\phi_2}) E_{-\alpha}
+(\mu e^{-\phi_1-\phi_2})g_cE_{-\alpha}g_c^{-1} \nonumber \\
&&+(\mu e^{-\phi_1+\phi_2}) E_{\alpha}
+ \left( \frac{2\pi}{k} \right) \left( \frac{\delta B}{\delta \phi_2} \right)
g_c h g_c^{-1}  
+\left( \frac{2\pi}{k} \right) \left( \frac{\delta B}{\delta \phi_1} \right)h =0.
\label{5.2} 
\end{eqnarray}
We try a Gauss decomposition to $g_c$:
\begin{equation}
g_c = e^{\lambda_1 E_{\alpha} } e^{\lambda_2 h} e^{\lambda_3 E_{-\alpha} },
\label{5.3}
\end{equation}
where $\lambda_1$, $\lambda_2$ and $\lambda_3$ are constants. Then, we find that
\begin{eqnarray}
&&g_cE_{-\alpha}g_c^{-1}= e^{-2\lambda_2}(
E_{-\alpha}+\lambda_1 h-\lambda_1^2 E_{\alpha}),\label{5.4} \\
&&g_chg_c^{-1}= h-2\lambda_1E_{\alpha}+2\lambda_3e^{-2\lambda_2}
(E_{-\alpha}+\lambda_1h-\lambda_1^2 E_{\alpha}) \ \ \hbox{and} \label{5.5} \\
&&g_cE_{\alpha}g_c^{-1}= e^{2\lambda_2} E_{\alpha}-\lambda_3
(h-2\lambda_1E_{\alpha}) -\lambda_3^2 e^{-2\lambda_2}
(E_{-\alpha}+\lambda_1h-\lambda_1^2E_{\alpha}).\label{5.6}
\end{eqnarray}

Now, as (\ref{5.2}) has terms proportional to $E_{\alpha}$,
$E_{-\alpha}$ and $h$, we have three independent equations. By
(\ref{2.42}), (\ref{2.45}) and (\ref{2.46}), we see that each
one of these equations has terms that are proportional to 
$e^{ \phi_1-\phi_2}$, $e^{ -\phi_1+\phi_2}$ and 
$e^{ -\phi_1-\phi_2}$. Thus, we have nine equations for the
three variables $\lambda_1$, $\lambda_2$ and $\lambda_3$. They 
can be solved and the result is
\begin{eqnarray}
\lambda_1= 2\lambda \ \ \hbox{and} \label{5.7} \\
\lambda_2=\lambda_3=0, \label{5.8}
\end{eqnarray}
where $\lambda$ is the constant introduced in the equations
(\ref{2.45}) and (\ref{2.46}).

\end{document}